\documentclass[sigconf,screen]{acmart}

\renewcommand\footnotetextcopyrightpermission[1]{} 

\AtBeginDocument{%
  \providecommand\BibTeX{{%
    \normalfont B\kern-0.5em{\scshape i\kern-0.25em b}\kern-0.8em\TeX}}}

\acmConference[CASES '22]{CASES '22:  International Conference on Compilers, Architecture, and Synthesis of Embedded Systems  (CASES)}{October 9--14, 2022}{Shanghai, China}

\usepackage{algorithm}
\usepackage{algpseudocode}
\makeatletter
\def\BState{\State\hskip-\ALG@thistlm}
\makeatother
\usepackage{amsmath}
\usepackage{multirow}
\usepackage{multicol}
\usepackage{array}
\usepackage{subfig}
\usepackage{caption}
\usepackage{balance}
\newcommand{\major}[1]{\textcolor{black}{#1}} 
\usepackage{fancyhdr}
\usepackage[absolute]{textpos}
\begin{document}

\title{MLGOPerf: An ML Guided Inliner to Optimize Performance}

\author{Amir H. Ashouri$^*$\quad Mostafa Elhoushi\quad Yuzhe Hua\quad  Xiang Wang\quad Muhammad Asif Manzoor\quad  Bryan Chan\quad Yaoqing Gao\vspace{1em}}
\thanks{*Corresponding author: amirh.ashouri@huawei.com}

\affiliation{%
  \country{Huawei Technologies, }
   \institution{Heterogeneous Compiler Lab \\Toronto, Canada}
}

\renewcommand{\shortauthors}{Ashouri et al.}

\begin{abstract}
For the past 25 years, we have witnessed an extensive application of Machine Learning to the Compiler space; the selection and the phase-ordering problem. However, limited works have been upstreamed into the state-of-the-art compilers, i.e., LLVM, to seamlessly integrate the former into the optimization pipeline of a compiler to be readily deployed by the user. MLGO was among the first of such projects and it only strives to reduce the code size of a binary with an ML-based Inliner using Reinforcement Learning. 
  
This paper presents MLGOPerf; the first end-to-end framework capable of optimizing performance using LLVM's ML-Inliner. It employs a secondary ML model to generate rewards used for training a retargeted Reinforcement learning agent, previously used as the primary model by MLGO. It does so by predicting the post-inlining  speedup of a function under analysis and it enables a fast training framework for the primary model which otherwise wouldn't be practical. The experimental results show MLGOPerf is able to gain up to 1.8\% and 2.2\% with respect to LLVM's optimization at O3 when trained for performance on SPEC CPU2006 and Cbench benchmarks, respectively. Furthermore, the proposed approach provides up to 26\% increased opportunities to autotune code regions for our benchmarks which can be translated into an additional 3.7\% speedup value.  
\end{abstract}

\ccsdesc[500]{Software and its engineering~Compilers}
\ccsdesc[500]{Computing methodologies~Machine learning}

\keywords{Compilers, Reinforcement Learning, Deep Learning, LLVM, Inlining}

\maketitle
\begin{textblock}{15}(2,1)
\noindent\Large The short version of this work is accepted at ACM/IEEE CASES'22 -- \url{https://esweek.org/cases}
\end{textblock}
\fancyhead{}

\section{Introduction}
\label{sec:Intro}

Compilers have gone a long way to emit efficient and high performance code given high-level programming languages. Breakthroughs in compiler technology are essential to making programming mainstream \cite{Hall2009}. State-of-the-art compiler frameworks such as LLVM and GCC are able to translate the high-level programming languages into an (almost) architecture-independent Intermediate Representation (IR) in which a sequence of optimization passes can be applied to optimize a code segment, iteratively. These optimization passes can be applied at the granularity of Function, Call Graph, Loop, or Module level \cite{lattner2008llvm}. Gradually and tentatively, compiler designers come up with a set of fixed-length optimization pipelines, known as standard optimization levels, i.e., \texttt{-O\{1,2,3,s,z\}}, that on average have shown benefits to optimizing performance (speed) or code size. For instance, LLVM's \texttt{O3} has around 160 passes in its pipeline, performing a number of optimizations at different levels of the granularity from which around 60 passes are unique and several sub-sequences of passes are repeated in the optimization sequence \cite{Ashouri2017micomp}. The idea behind producing such a standardized optimization sequence is to enable users with an option that performs on average good \cite{georgiou2018less}.

In the past couple of decades, several attempts have been made to specialize the standard optimization levels given a code segment of interest as they are not always the optimal solution \cite{Agakov2006,chen2012deconstructing,Park2013,Ashouri2016Cobayn,Ashouri2017micomp}. These approaches rely on leveraging applications of Machine Learning (ML) and an automatic tuning methodology, possibly using an autotuner \cite{ansel2014opentuner,huawei2019autotuner}, to speed up the search in the vast compiler optimization space \cite{ashouri2018survey}. Although in many cases researchers have achieved meaningful results, these approaches are often limited by the use of external/third-party components outside a compiler framework to derive intelligence for it. Fortunately, we have been witnessing more and more proposed works, tackling the above problems by adding end-to-end frameworks \cite{cummins2017deep,haj2020neurovectorizer,das2020GraghColoringAMD,rotem2021profilecummins,NeuralInstructionCombinerIntel2022}, however, to the best of our knowledge, there has been only a single recent work, namely MLGO \cite{trofin2021mlgo}, which does so by refactoring an optimization pass, i.e, function inlining, to bring an ML-based guidance into the optimization pass, as a standalone component. MLGO is also upstreamed into LLVM trunk \footnote{https://lists.llvm.org/pipermail/llvm-dev/2020-April/140763.html}.

Function Inlining is one of the fundamental compiler optimizations used by many state-of-the-art compiler frameworks; when applied to a function, it examines each call site within the function and decides whether or not to inline the function body of the callee into the caller.
Not only does inlining eliminate the function call overhead at the call site, it also expands the scope of intra-procedural analyses of subsequent passes and enables additional optimizations \cite{scheifler1977inline,theodoridis2022inlining}. When performed incorrectly, however, inlining can cause code size increase, which can lead to performance loss due to instruction cache misses. Therefore, an ideal inlining heuristic must either improve performance without incurring unacceptable code bloat, or reduce code size while avoiding substantial performance loss. Constructing an ideal inlining heuristic has been shown to be a complex problem and the rate at which the number of choices would grow is known to be at least at an exponential increase \cite{ashouri2018survey,theodoridis2022inlining}.

Presently, MLGO provides code size reduction to the LLVM's inline optimization and we based our proposed work on adapting a framework which provides performance optimization for it. Our work, MLGOPerf, proposes the aforementioned contribution to the community, by leveraging two ML models to optimize inline optimization for speed. MLGOPerf, increases the tunable code-regions subsequent to the inline pass under \texttt{O3} and we show in the experimental results that it can provide an added speedup value with respect to the MLGO's and the vanilla inliner. MLGOPerf proposes the following contributions:

\begin{enumerate}
    \item We propose an ML model, namely \texttt{IR2Perf}, to predict the post-inlining function speedup of a caller with respect to its baseline. IR2Perf leverages a number of handcrafted features from LLVM IR and it correlates the speedup outcome to the changes in IR as a result of function inlining. The model enables us to rapidly generate \texttt{Rewards} needed to train the existing Reinforcement Learning (RL) agent used for LLVM's \texttt{ML-Inliner} infrastructure without the need to execute each program thousands of times.
    \item We add an extra couple of features to better define the behavior of callee functions and to boost the accuracy of the existing LLVM ML-Inliner RL agent.
    \item At model training, we utilize the newly generated rewards from IR2Perf and revised the RL agent to optimize for performance rather than code size. Finally, at model deployment, we provide a pretrained model to be built with LLVM CMake system and used with LLVM's ML-Inliner for inference without the need for IR2Perf. 
\end{enumerate}

The rest of the paper organizes as follow. Section \ref{sec:related} discusses the state-of-the-art. Section \ref{sec:proposed} provides details on our proposed method and how the two ML models interact with each other in Sections \ref{sec:proposed:ir2perf_rl} and \ref{sec:proposed:phases}, respectively. Section \ref{sec:res} showcases the experimental results. Finally, in Section \ref{sec:discussion} we reveal the current shortcomings, challenges, and propose some future work. 

\section{Related Work}
\label{sec:related}

Compiler autotuning has become an extensively researched area in the past two decades \cite{ashouri2018survey}, more so with the introduction of the application of Machine Learning (ML) in a number of early approaches, i.e, optimization space reduction \cite{Cooper1999}, estimating unrolling factor \cite{Koseki1997}, scheduling \cite{cavazos1998}, etc. However, the space has gradually shifted towards tackling two major problems, both of which remains open problems, namely (1) the optimization selection problem \cite{bodin1998iterative, Agakov2006,fursin2009collective,Park2013,Ashouri2016Cobayn} and, (2) the phase ordering problem  \cite{Kulkarni2012,Ashouri2017micomp,nobrePhase2018,cummins2021compilergym,wang2022RLcompilerGym}. 

The bulk of the approaches are tackling the above problems as a black-box optimization; by means of an autotuning \cite{Fursin2011, chen2012deconstructing, Ashouri2013VLIW, ansel2014opentuner, huawei2019autotuner} strategy paired with an ML model, which derives an iterative compilation \cite{bodin1998iterative,Agakov2006,Hoste2008,Tiwari2009,Park2012,Ashouri2016Cobayn,Ashouri2017micomp,Ashouri2018book} methodology to speed up the search. 

Recently, we have also witnessed the applications of Deep Learning in the mix. For instance, using Deep Reinforcement Learning \cite{sutton2018reinforcement} has become a popular method. On the optimization side, authors tackle the problem of finding the optimal vectorization \cite{haj2020neurovectorizer}, providing a high-level optimization environment for compiler research \cite{cummins2021compilergym}, and an autotuner \cite{park2022srtuner}. Additionally, by leveraging Natural Language Processing (NLP) \cite{young2018NLP-RL}, one can generate an embedding or simply a characterization of the source-level software using representative features \cite{ben2018ins2vec,alon2019code2vec,cummins2021programl}.  However, there have been a number of works addressing the aforementioned problems with an end-to-end framework and as such, they are tailored towards bringing the autotuning into the compiler \cite{Fursin2008,cummins2017deep, haj2020neurovectorizer,cummins2021compilergym,rotem2021profilecummins,das2020GraghColoringAMD,NeuralInstructionCombinerIntel2022}. These methods still leverage a high-level optimization engine which wraps around the identification of beneficial passes to apply given a segment of code and none has managed to propose a standalone optimization pass capable of deriving decisions by means of inference from an ML model built-into the compiler.

Rotem and Cummins \cite{rotem2021profilecummins} propose an ML framework, leveraging Decision Trees (DS), to provide profile-guided Optimization (PGO) to branch probabilities. The authors use handcrafted features for both basic blocks and branches to characterize a code segment and employ XGBoost \cite{chen2016xgboost} library to generate their DS. This work is relevant to ours only because it does the aforementioned methodology built into LLVM and once trained, it can provide PGO inferences without the need to actually executing a number of iterations of PGO profiling. 

Phothilimthana et al. \cite{mangpo2021} propose a two-level autotuner to tune graph-level and subgraph-level optimizations across multiple compilation stages.  The authors provide a joint optimization methodology for a number of tensor parameters and operations mostly leveraged in production ML compilers, including tile size, tensor layouts, operator fusion, and code generation. This work is orthogonal to our proposed work.

Trofin et al. \cite{trofin2021mlgo} propose MLGO which was the first of a kind to provide an ML-guided optimization for a pass, i.e., inline optimization \footnote{The repository has since provided an added option for register allocation.}. The work has been upstreamed into LLVM and provides guidance to the inline optimization as to whether or not a call site should be inlined, provided that the generated code size is minimized. Our proposed work, MLGOPerf, is similar to MLGO in the sense that we leverage the inlining infrastructure already upstreamed in LLVM repository, however, we adapt a framework to derive decisions to optimize performance rather than a reduction in code size. 

Optimizing the execution-time performance of compiler passes is an inherently more difficult problem and we attempt to do so by proposing a second ML model, i.e., \texttt{IR2Perf}, which is able to predict the speedup of a function after it was passed through the function inlining optimization. Leveraging IR2Perf, we generate a fitness function, namely, Rewards to train an RL agent that learns the complex behavior of providing an inlining decision given a  call site. Recent studies have suggested using a semi-supervised learning \cite{konyushkova2020semisupervisedRL} and using loss as self-supervision \cite{shelhamer2016lossSupvervisedRL} to guide the learning of an RL agent with success. We believe, a supervised reward learning process has merits in helping the ML-Inliner agent to learn its optimal policy. Once trained, MLGOPerf is able to provide standalone predictions to LLVM's ML-Inliner and to optimize segments of the code that are (1) running faster and additionally, (2) contain an enhanced number of inter-procedural opportunities down the pipeline of LLVM's O3 which may translate into a faster code.  

\section{Proposed Methodology}
\label{sec:proposed}

Recently, Upstream LLVM Inliner has benefited from MLGO \cite{trofin2021mlgo} for which the user can leverage an ML-Inliner pass to derive Inlining decisions for  call sites. The current ML-Inliner approach is targeted towards code size reduction and, \major{although function inlining is not the first candidate among compiler passes for directly increasing the performance of running codes} \cite{scheifler1977inline,prokopec2019inline,theodoridis2022inlining}, we believe there is merit in designing a flow in which the ML-Inliner targets a performance gain when decides whether or not to inline a  call site. An added benefit of this approach is the new opportunities proper function inlining can provide for subsequent compiler passes within the \texttt{O3} pipeline. 

\begin{figure}[t]
\centering
\includegraphics[width=.5\textwidth]{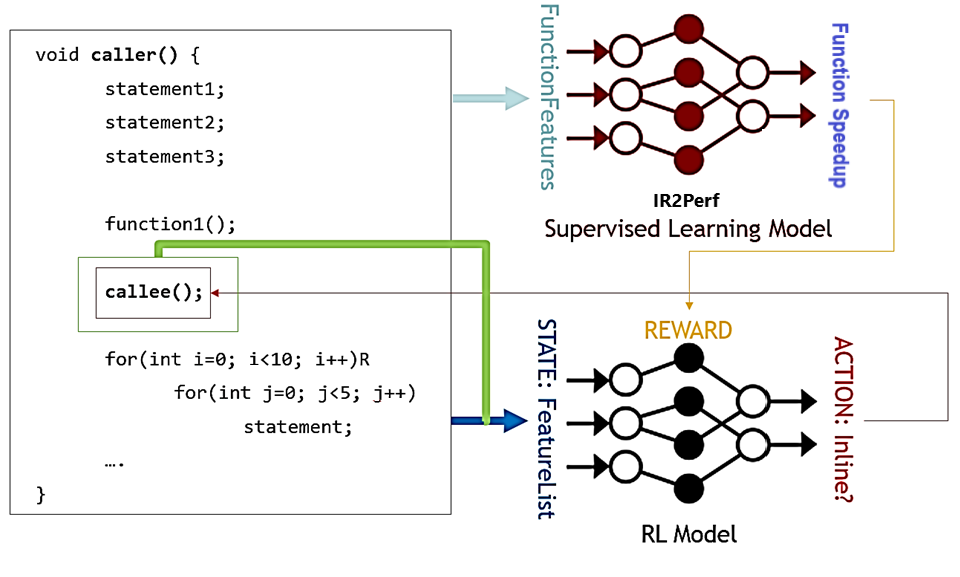}
\caption{MLGOPerf Highlevel Flow} 
\label{fig:mlgoModels}
\end{figure}

There are three major challenges and limitations to achieving such behavior: (1) The current Reinforcement Learning (RL) infrastructure designed to reproduce MLGO \footnote{https://github.com/google/ml-compiler-opt} only tackles the reduction of  the code size of a function and we need to adapt the methodology to optimize performance instead. (2) It relies on a set of rewards to guide the training of its policy trajectory\cite{schulman2017ppo}. (3) training such RL agent requires thousands of evaluations or observations and at each iteration, a reward value is needed to learn the profitability of the action. Therefore, using the actual execution time cannot be a practical solution; this is especially a problem for real-world applications, e.g., SPEC, which takes tens of minutes per round of execution. To this end, we design a second ML model, i.e., IR2Perf, to help alleviate the above challenges. Figure \ref{fig:mlgoModels} depicts the systematic view of the interaction between the two ML models in our work.

\subsection{IR2Perf to the Rescue}
\label{sec:proposed:ir2perf_rl}

At the granularity of  call sites, ML-Inliner decides whether or not to inline the function body of the \texttt{callee()} into the \texttt{caller()}'s. It does so by using the existing Inference path under \texttt{MLInlineAdvisor: InlineAdvisor}. It collects a number of features, including CalleeBasicBlockCount, CallSiteHeight, NodeCount, etc., to provide an inference to the advisor. However, in order to train such a model, one requires rewards to be fed, at each iteration of its training pipeline, to guide the trajectory of its policy decisions when it takes an action\cite{trofin2021mlgo}. We provide IR2Perf model at the granularity of Function level for our \texttt{caller()} to predict the speedup of the whole function as a proxy to generate rewards for the RL model. 

\subsection{MLGOPerf Phases}
\label{sec:proposed:phases}

\begin{figure*}[p!]
\centering
\includegraphics[width=.85\textwidth]{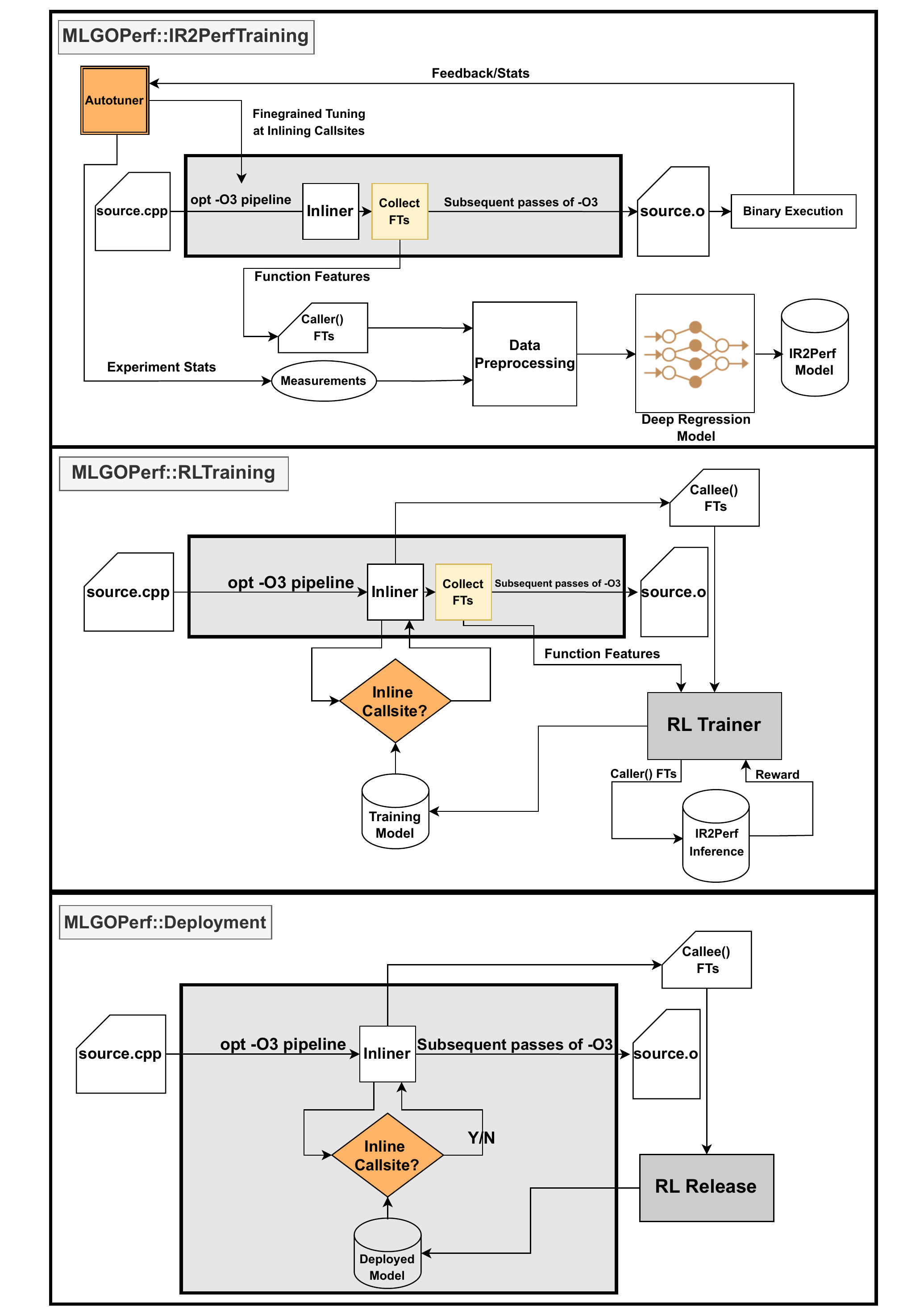}
\caption{MLGOPerf Three Phases} 
\label{fig:mlgophases}
\end{figure*}

As a result of leveraging two ML models, our proposed work is a multi-phased methodology and thus, in this section we provide insights to better demonstrate the functionalities of our approach. Table \ref{tab:phases} showcase the specific interaction between the two models within the three phases. Furthermore, Figure \ref{fig:mlgophases} depicts the three phases in higher detail. We discuss the three phases in the following subsections.

\begin{table}[t!]
    \centering
        \caption{MLGOPerf Phases}
    \begin{tabular}{ |l|c|c|}
        \hline
        Phases & IR2Perf & RL Model \\ \hline
        IR2Perf Training & Training & -
        \\ \hline
         RL Model Training & Inference &  Training
        \\ \hline
        MLGOPerf Deployment & - & Inference
        \\ \hline
    \end{tabular}
    \label{tab:phases}
\end{table}

\subsubsection{IR2Perf Training}
\label{sec:proposed:phases:ir2perfTraining}

In this phase, we design an autotuning methodology that controls the Inlining decisions made at the granularity of the  call sites of a function. We do so by leveraging an autotuner \cite{huawei2019autotuner}, an OpenTuner \cite{ansel2014opentuner} derivative. This allows us to generate meaningful training data that captures the behavior of function inlining when it decides to inline a  call site into its respective caller together with its \texttt{Function} and \texttt{Module} execution times. Additionally, we develop a number of relevant function features, as a pass, and register it subsequent to the Inline optimization into the \texttt{O3} pipeline to collect our training data. 

Let $\omega$ be a characterization vector of features of a function having at minimum one  call site. This vector represents $l$ variables to account for the intermediate representation of the function when collected post-inlining. We feed the collected features together with the measured execution time and the corresponding inlining configuration into our ML model. IR2Perf is a Deep Regression model, when trained, designed to predict the relative speedup of a function under analysis wrt \texttt{O3} when an inlining configuration was applied.  

\subsubsection{RL Model Training}
\label{sec:rl_model_training}

\major{Reinforcement Learning (RL)} is a class of Machine Learning which tries to find an optimal policy for a Markov Decision Process (MDP) that is defined by the tuple of $<S, A, T, R>$ where $S$ is the state space, $A$ represents action space, $R$ is the reward function that an agent receives by doing action a from state s to s', and $T$ is the transition probability at time t from state s to s': $T_{a}(s,s')=Pr(s_{t+1}=s'|s_t=s,a_t=a)$ \cite{sutton2018reinforcement}. The goal of RL training is for its agent to learn an optimal policy to maximize its reward function. In this work, we formulate the tuple as follows:

\begin{enumerate}
    \item State $S$: Current visiting  call site
    \item Action $A$: Defines a bool variable whether or not to inline the  call site
    \item Transition $T$: A deterministic function in our context which updates the call graph upon taking an action over the current  call site and switching to visit the next  call site 
    \item Reward $R$: In this work, it is defined as the function execution speedup with respect to the baseline 
\end{enumerate}

As stated in MLGO \cite{trofin2021mlgo}, training the original method for Speed has difficulties for runtime measurement of a large body of benchmarks and a more complex problem with respect to the training for code size. Unlike a code size reduction objective, we cannot decouple performance by means of a Commutative and Associative properties in an optimization exploration strategy. Therefore, we plan to address this issue by leveraging IR2Perf. There are a number of methods to train such RL agent which uses Proximal Policy Optimization (PPO) \cite{schulman2017ppo} algorithm. Algorithm \ref{algo:trainingRL} provides an insight into the training flow for which we update the policy of our RL agent using Equation \ref{eq:RL}. Similar to MLGO, we rely on total reward as the summation of all the rewards generated using IR2Perf. It is worth noting that using IR2Perf, we could also generate partial rewards per function at each time $t$, but we did not follow this method. 

\begin{algorithm}[!t]
\caption{Training Inliner RL Model using IR2Perf}
\label{algo:trainingRL}
\begin{algorithmic}[1]
\Procedure{FunctionSpeedup}{} \Comment{IR2Perf Inference }
      \For{\texttt{Function f in Module}}
        \State $FTs \gets \textit{getFunctionFeatures()}$
        \State $funcReward \gets infer(FTs) $
        \State $totalReward \gets append(funcReward)$
      \EndFor
      
\State \Return \textit{totalReward}
\EndProcedure
\State
\Procedure{CallsiteInline}{} \Comment{RL Model Training}
\State initialize policy $\pi_{\gamma}$ \major{randomly}
      \For{\texttt{iteration i in Training}}
      \State $s_{i} \gets Sample_{N(0,I)}(TrainingData)$
      \State Compile and Get IR with policy $\pi_{\gamma+\sigma s_{i}}$
      \State $R \gets FunctionSpeedup(Module)$
      \State Update policy $\gamma$ for using Equation \ref{eq:RL}
      \EndFor
\EndProcedure
\end{algorithmic}
\end{algorithm}

\begin{equation}
    \label{eq:RL}
    \gamma = \gamma + \alpha\frac{1}{n} \Sigma^n_{i=1}\{\Sigma^{t_{n}}_{t=0} R_{total_{i}} \nabla_{\gamma} log \pi_{\gamma}(a_{i,t}|s_{i,t}) \}
\end{equation}

\noindent where $\alpha$  is the learning rate at which the policy $\gamma$ is updated. The RL agent tries to maximize the total reward of its policy when receiving from IR2Perf when it applies an action $(a_{i,t}|s_{i,t})$.

\subsubsection{MLGOPerf Deployment}

After our RL model was trained for sufficient iterations \cite{machado2018rlTraining}. we unplug IR2Perf and build LLVM with the pretrained RL model. The point at which we stop the training can be determined by occasional evaluation of the performance of the RL model, observing the trajectory of the loss function, or, by means of a training budget constraint, i.e., number of iterations or allocated time. This step is similar to the \texttt{Release Mode} step in MLGO \cite{trofin2021mlgo}.

\section{Experimental Results} 
\label{sec:res}

In this section, we provide the experimental results of our proposed work. From the perspective of Compiler Engineers and in order to fine-tune the quality of the predictions, one has to reproduce the first two phases stated in Section \ref{sec:proposed:phases} and, from the user's perspective, the third and the final phase is of relevance where we deploy the pretrained model and build it together with LLVM infrastructure. At deployment, there is no need to leverage neither IR2Perf nor RL Model since IR2Perf is no longer needed to generate rewards, and also, the RL model can be compiled and built AOT (Ahead Of Time) using \texttt{saved\_model\_cli} tool \footnote{https://developers.googleblog.com/2017/03/xla-tensorflow-compiled.html}. 


\subsection{IR2Perf Model}
\label{sec:res:ir2perf}

As we discussed in Section \ref{sec:proposed:phases:ir2perfTraining}, we develop a pass and register it subsequent to the Inline optimization in \texttt{O3} pipeline to collect the handcrafted features corresponding to a function of interest. In recent years, there have been a number of approaches \cite{ben2018ins2vec,alon2019code2vec,cummins2021programl} proposed to provide an embedding of the Intermediate Representation (IR) of the program for which we could have used but instead, in this work, we decided to directly characterize the space using static features of IR and we show them in Table \ref{tab:ir2perfFTs}. \major{The overhead caused by collecting these features is negligible.}

We collect a total of 370K different inlining configurations from 10 SPEC CPU2006 benchmarks using our autotuner for which we tuned the inlining parameter at the granularity of  call site. For practicality, we used the \texttt{train} dataset for SPEC 2006 as opposed to the \texttt{ref} dataset; this limits the runtime of our training samples to be within a minute each on average with a few, i.e., 403.gcc, 462.libquantum, etc., being on the shorter execution time range. Note that the evaluations in Section \ref{sec:res:mlgoperfDeployment} are still performed with the \texttt{ref} dataset. \major{This process takes around 5 days. Execution time of a function call can be too short to allow for a reliable measurement and to this end, we exclude speedup values higher than our exclusion threshold. In this work, we use 3$\times$ as the exclusion threshold hyperparameter. Additionally, we filter out recorded function overheads of less than than 1\% by Perf to avoid a noisy estimate in our training data.}

One challenge here is to measure the exact CPU time the program takes on every call of a certain function. We estimate this with the following: 

\begin{equation}
    \label{eq:IR2Perf_runtime_calculation}
    Func_{runtime} = \frac{(\text{Total Program Runtime)} * T_{Func}}{N_{Func}}
\end{equation}

\noindent where $T_{Func}$ is the percentage of time the program spent on this function and $N_{Func}$ the number of times this function was called during the execution.

Subsequently, we define the function speedup as the true label for IR2Perf model as: 

\begin{equation}
    \label{eq:IR2Perf_speedup_calculation}
    Func_{speedup} = \frac{Func_{runtime(Base)}}{Func_{runtime(X)}}
\end{equation}

\noindent where $Func_{runtime(Base)}$ is the execution time of the function with no inlining configuration and $Func_{runtime(X)}$, the same function having an inlining configuration of X in the set of all inlining configurations generated by our autotuner. Total program runtime and the $T_{Func}$ are collected with Linux \texttt{perf} and $N_{Func}$ is collected with profiling instrumentation and \texttt{llvm-profdata} tool. In this work, all runtime measurements are done single-threaded, and we use \texttt{numactl} tool to bind the workloads to a unique CPU core. \major{After each measurement run, we flush the system page cache to avoid any perturbance  into the collection of our training data.} 

The goal is to include diverse and meaningful inlining configurations. Each tuning iteration provides us, depending on the number of  call sites, a number of data points which can be added to our training data. Many of the function characteristics are correlated to one another in complex ways and also, with the target metric, i.e., function speedup. Thus, we perform a preprocessing stage to better represent the data for training our model. It is well-known that the optimal function inlining problem has an exponential increase in optimization space \cite{theodoridis2022inlining,ashouri2018survey}. Therefore, generating sufficient training data which captures as many permutations of combinations of inlinig parameters still remains a pain point. We apply a dimension reduction of features space since it is experimentally observed as the feature space increases, so does the average distance between points. Multidimensional datasets, similar to ours, are subject to rarity \cite{james2013pca} and to this end, we employ Principal Component Analysis (PCA) setting PC to 7 to reduce the dimensionality of our feature space while preserving the information of our training data with minimal loss \cite{deco1996informationPCA}.

\major{The preprocessing is done as the following: (0) We compute the global and function speedups of each inlining configuration against its baseline, i.e, \texttt{O3} (1) We then remove redundant data points from the training data at this stage and normalize the dataset, followed by (3) applying PCA and additionally, as a standard practice, we store the objects of our feature scaling and PCA process to transform our test data into the same scale determined by the training set used to train IR2Perf.}

\begin{table}[t!]
    \centering
        \caption{IR2Perf Model Features}
    \begin{tabular}{|>{\centering\arraybackslash}m{0.4cm}|l|c|}
    \hline
        No & Static Feature Name  \\ \hline
        1     &  InstructionPerBlock     \\ \hline
        2     &  SuccessorPerBlock    \\ \hline
        3     &  AvgNestedLoopLevel   \\ \hline
        4     &  InstrPerLoop   \\ \hline
        5     &  BlockWithMultipleSuccecorsPerLoop    \\ \hline
        6     &  CallsNo   \\ \hline
        7     &  IsLocal   \\ \hline 
        8     &  MaxLoopDepth   \\ \hline
        9     &  MaxDomTreeLevel    \\ \hline
        10     &  CallerHeight   \\ \hline
        11     &  CallUsage   \\ \hline
        12     &  IsRecursive    \\ \hline
        13     &  NumCallsiteInLoop   \\ \hline
        14     &  EntryBlockFreq   \\ \hline
        15     &  MaxCallsiteBlockFreq   \\ \hline
        16     &  NoOfInstructions='Ret' \\ \hline
        17     &  NoOfInstructions='fmul' \\ \hline
        18     &  NoOfInstructions='fdiv' \\ \hline
        19     &  NoOfInstructions='fadd' \\ \hline
        20     &  NoOfInstructions='fsub' \\ \hline
    \end{tabular}
    \label{tab:ir2perfFTs}
\end{table}

IR2Perf is a Regression model for which there are four linear fully-connected layers followed by an activation function. Here, we use \texttt{Leaky RelU} rather than \texttt{ReLU} as we experimentally observe a slight benefit in the accuracy of the model. 

\begin{table}[t]
    \centering
        \caption{IR2Perf Model Architecture}
    \begin{tabular}{|>{\centering\arraybackslash}m{0.8cm}|l|c|}
    \hline
        Layer No & Layer (type) & Output Shape \\ \hline
       \multirow{2}{*}{1} &   Linear-FC1   & [-1, 1, 128]  \\ 
          & Leaky\_Relu-1   & [-1, 1, 128]  \\ \hline
        \multirow{2}{*}{2}   & Linear-FC2   & [-1, 1, 256]  \\ 
          & Leaky\_Relu-2   & [-1, 1, 256]  \\ \hline
        \multirow{2}{*}{3}   & Linear-FC3   & [-1, 1, 32]  \\ 
          & Leaky\_Relu-3   & [-1, 1, 32]  \\ \hline
        4   & Linear-FC4   & [-1, 1, 1]  \\ \hline
    \end{tabular}
    \label{tab:ir2perfarch}
\end{table}

\subsection{IR2Perf Accuracy}
\label{sec:res:ir2perfAccuracy}

\begin{figure*}[t!]
\centering
\subfloat[Training Loss  \\ (First epoch)]{
    \label{fig:ir2perf_lossPlot}
            \resizebox{.2\textwidth}{.17\textwidth}{\includegraphics{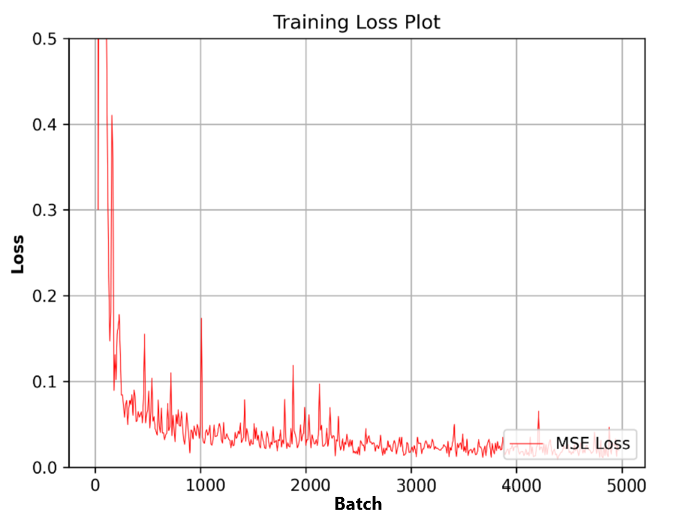}}}
\captionsetup[subfloat]{margin=0.75cm}
\subfloat[Training Accuracy \\ (\texttt{401.bzip2} was held out)]{
    \label{fig:ir2perf_traininigPlot}
            \resizebox{.25\textwidth}{!}{\includegraphics{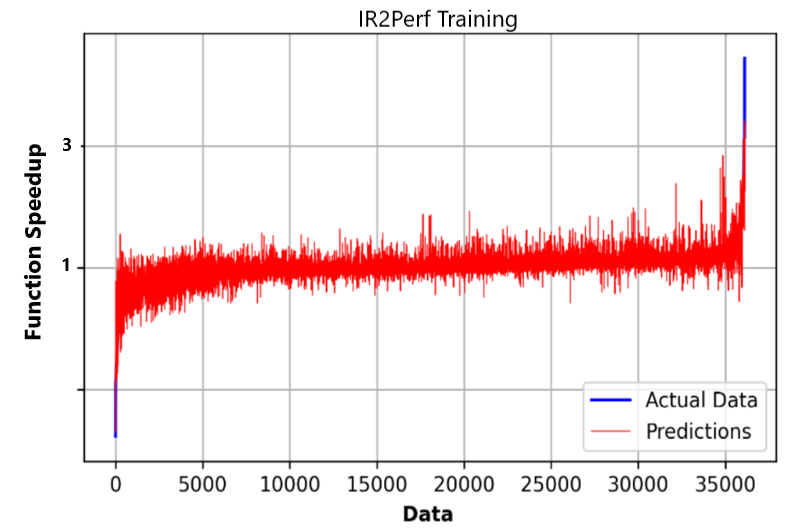}}}
\subfloat[Prediction Accuracy of \\ \texttt{401.bzip2} hold-out on \ref{fig:ir2perf_traininigPlot}]{
    \label{fig:ir2perf_predictionPlot_bzip}
            \resizebox{.25\textwidth}{!}{\includegraphics{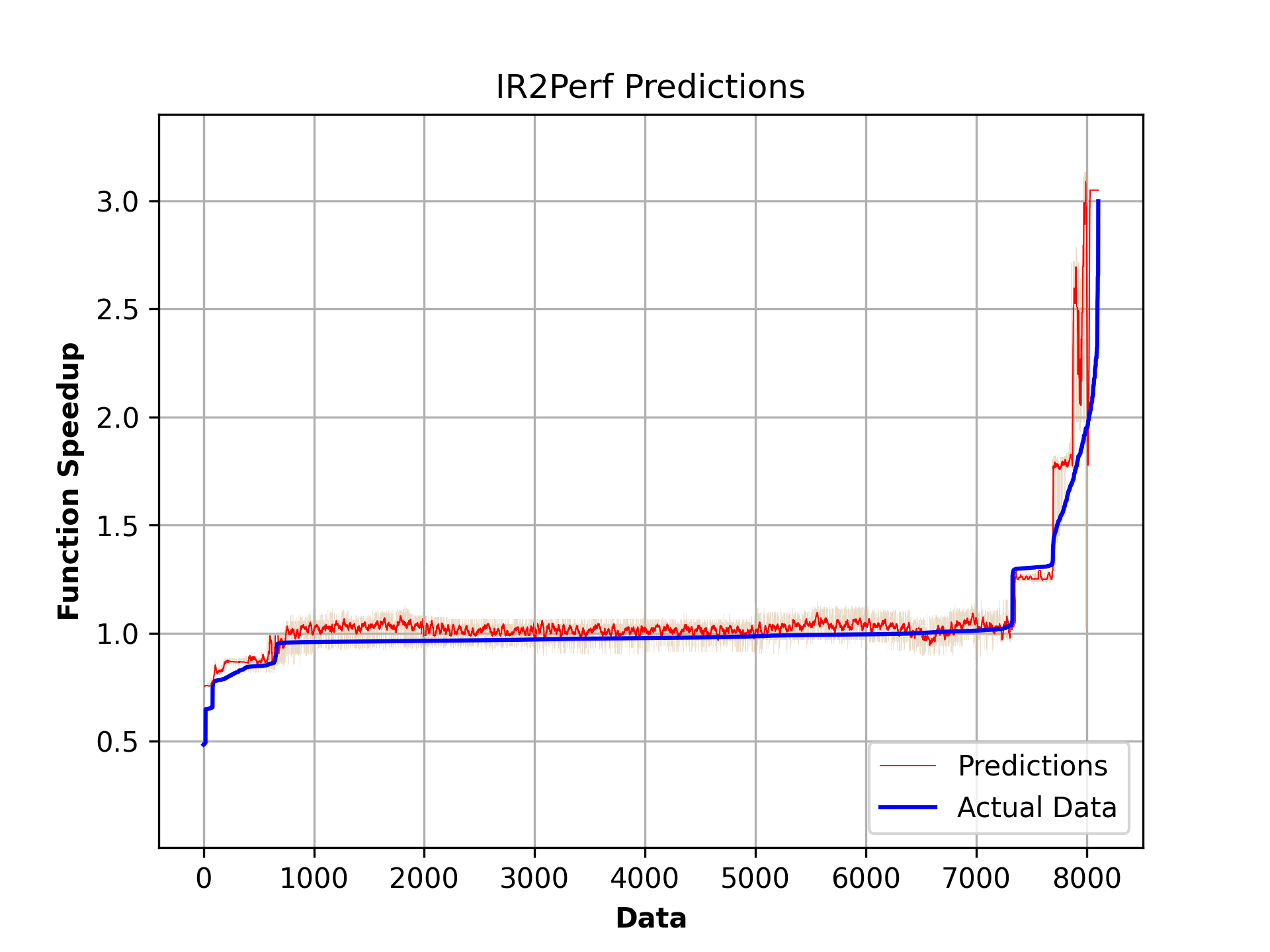}}}
\subfloat[Prediction Accuracy of \\ \texttt{464.h264ref} as hold-out (training figure not shown)]{
    \label{fig:ir2perf_predictionPlot_h264}
            \resizebox{.25\textwidth}{!}{\includegraphics{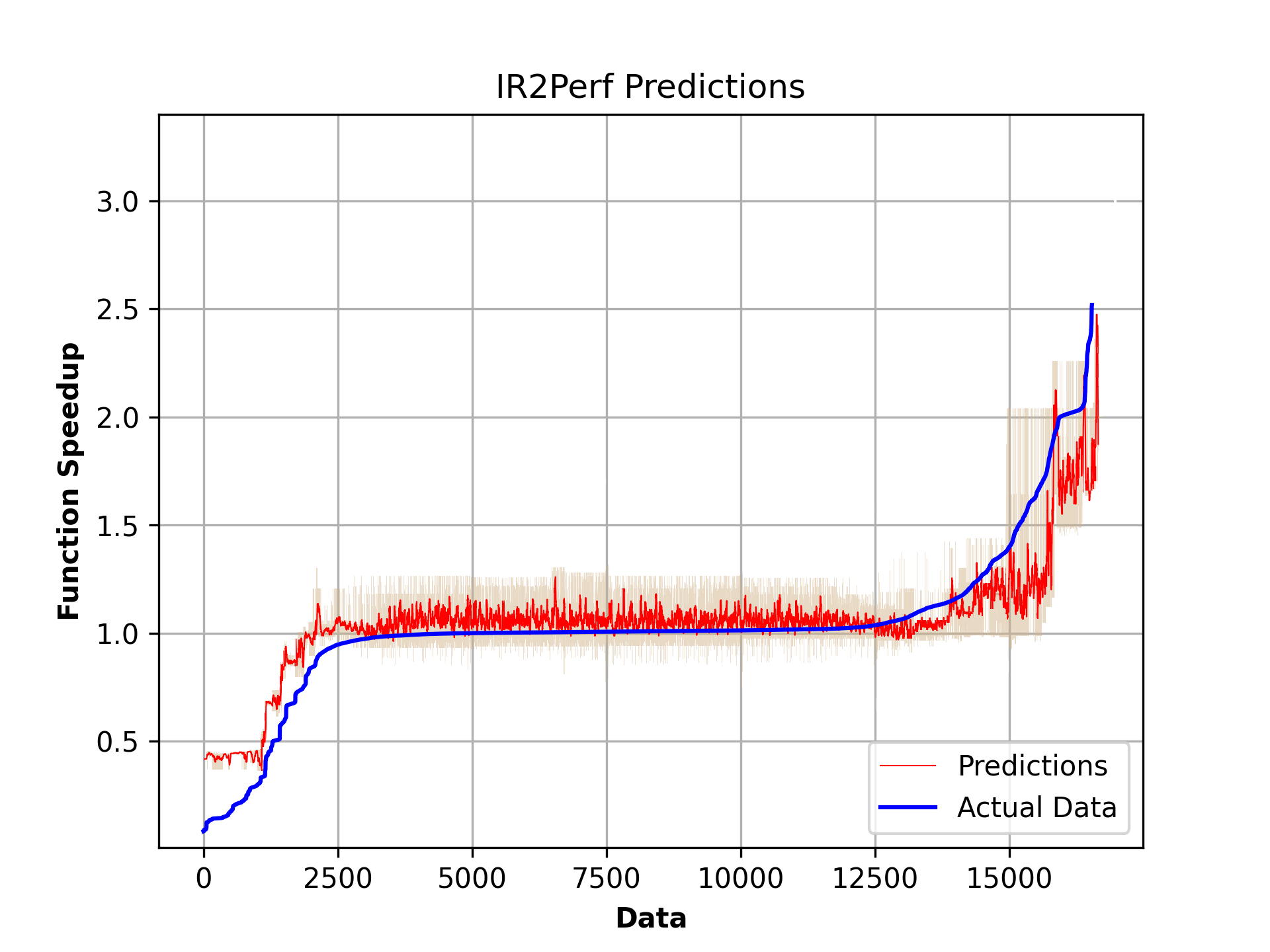}}}
\caption{IR2Perf Model Evaluation}
\label{fig:if2perfTrainingResults}
\end{figure*}

The training is done using leave-one-out cross-validation for which we exclude one benchmark from the training data and use it later as our test data. This ensures no data leakage occurs.  Table \ref{tab:ir2perfAccuracy} represents the accuracy of IR2Perf for each benchmark. In this work, we use Mean Squared Error (MSE) loss function and we observe that on average, IR2Perf achieved an error rate of 12.8\% among the SPEC CPU2006 benchmarks with a best 9.1\% on \texttt{401.bzip2}. Figure \ref{fig:if2perfTrainingResults} depicts a number of IR2Perf ML accuracy evaluations. Specifically, Figure \ref{fig:ir2perf_lossPlot} represents the loss function of the first epoch of our training. As can be seen, the model converges rapidly with a few iterations of batches. Figure \ref{fig:if2perfTrainingResults} showcases the training graph when \texttt{401.bzip2} was excluded from the training data. Subsequently, Figure \ref{fig:ir2perf_predictionPlot_bzip} shows the prediction accuracy after the model was tested on \texttt{401.bzip2}. Finally, Figure \ref{fig:ir2perf_predictionPlot_h264} depicts the prediction accuracy of IR2Perf, i.e., 19.1\% when \texttt{464.h264ref} was excluded from the training set (training figure is not shown here for conciseness). \major{The prediction accuracy of IR2Perf is established by the process mentioned above and we use the pretrained model with the lowest error-rate found by means of cross-validation. We chose SPEC as the main suite for data collection as it provides real-world complex function-level code segments which can benefit IR2Perf to characterize the behaviour of inlining.}

\begin{table}[t!]
    \centering
        \caption{IR2Perf Cross Validation Accuracy}
    \begin{tabular}{|c|l|>{\centering\arraybackslash}m{1.5cm}|}
    \hline
        NO. & Benchmark & Prediction Error (MSE)   \\ \hline
        1 & 401\_bzip2 & \textbf{9.1}\%     \\ \hline
        2 & 456\_hmmer &  9.5\%     \\ \hline
        3 & 462\_libquantum &  15.7\%    \\ \hline
        4 & 464\_h264ref &    \textbf{19.1}\%  \\ \hline
        5 & 445\_gobmk &     17.5\% \\ \hline
        6 & 470\_lbm &      9.8\%\\ \hline
        7 & 458\_sjeng &     13.5\% \\ \hline
        8 & 429\_mcf &     12.2\% \\ \hline
        9 & 433\_milc &     13.9\% \\ \hline
        10 & 482\_sphinx3 &   11.2\%   \\ \hline
       \multicolumn{2}{|l|}{Geometric Mean} & 12.8\% \\ \hline
    \end{tabular}
    \label{tab:ir2perfAccuracy}
\end{table}

\subsection{MLGOPerf Training}
\label{sec:res:mlgoperfTraining}  
 
Once the efficient accuracy of IR2Perf is established, we infer from it to generate rewards for our RL agent trainer. In this stage, i.e., phase 2 of Figure \ref{fig:mlgophases}, we deploy IR2Perf into \texttt{MLGOPerf::RLTraining} pipeline, and given an incoming function in a module, we collect the caller's function feature to be fed as input to IR2Perf. This will guide the RL agent's policy as to how well an inlining configuration potentially performs given its callee function representation which we also collect and feed into the trainer. In the original version of MLGO \cite{trofin2021mlgo}, authors propose 11 callee features to be used with the RL trainer \footnote{Specifically, we use LLVM 12 as at commit 1dad9d42 and MLGO Github repository as at commit dac1b149}, we add two more relevant features we believe have added representational value, namely (1) Block Frequency and (2) Loop Level. \major{Our early results show benefits to the final MLGOPerf model when we employ our two handcrafted features to the mix by around 1\%.} These two will be used together with the previously introduced 11 features and they characterize the caller function for the RL agent.

We train the RL agent using the default hyperparameters mentioned in the original MLGO work, except enabling rewards to be normalized as suggested by \cite{schulman2017ppo}. Rapidly generating rewards using IR2Perf, training the RL agent takes around 4 days using an NVIDIA Tesla P100 PCIe 12GB and an Intel(R) Xeon(R) CPU E7-8890 v4 @ 2.20GHz on Linux 18.04.1 LTS when trained on module samples taken from SPEC CPU2006. Note that IR2Perf reward generator is not limited to any benchmark it was trained on or cross-validated with, as it can liberally infer the function speedup of any unseen function when only fed the IR features defined under Table \ref{tab:ir2perfFTs} and thus, we successfully eliminate the need to collect any runtime metrics, i.e, execution time, and to largely accelerate the training process.

\subsection{MLGOPerf Deployment}
\label{sec:res:mlgoperfDeployment}   

We deploy the MLGOPerf under the third and final phase of our proposed work. Similar to MLGO, the deployed model when compiled AOT is emitted as a library which can be invoked when we feed as input the 13 callee features of any unseen applications. Note that at this phase, we no longer need to use IR2Perf as it has already done its job to help train the RL agent. \major{The overhead of collecting these callee features is minimal and MLInliner framework can leverage the decision made by the RL model to decide whether or not to inline a  call site to optimize the performance of an application under analysis}. Similar to Section \ref{sec:res:ir2perf}, all run time measurements are done single-threaded, and we use \texttt{numactl} tool to bind the workloads to a unique CPU core and we made sure we flush the system page cache after every measurement to avoid any protuberance into the training data. The experiments are measured on an ARMv8.2-A (Kunpeng 920) architecture @ 2.6GHz on Linux. We run each benchmark five times, having flushed the system page cache after every run, and we use a trimmed mean method to drop the minimum and the maximum measurements and perform an average of the remaining three measurements. Additionally, we collect the variance between the measurements and we make sure to repeat the measurements for every instance of evaluation having a variance of more than 2\%.

\subsection{Performance Improvement}
\label{sec:res:standardOpt}

Table \ref{tab:mlgoperf_results} shows our experimental results using SPEC CPU2006 with the \texttt{ref} dataset (Table \ref{tab:mlgoperf_results_spec}) and Cbench (Table \ref{tab:mlgoperf_results_cbench}). As can be seen, MLGOPerf on average achieves 1.8\% and 2.2\% speedup against LLVM's O3 on SPEC CPU2006 and Cbench, respectively. Additionally, we compare the performance of MLGOPerf with respect to MLGO and on average it outperforms MLGO by 3.1\% and 4.1\% on SPEC2006 CPU and Cbench, respectively.  The average variance between the runs is measured to be around 0.3\% to 0.4\% on SPEC and 0.43\% to 0.86\% on Cbench. As expected, there is a slight increase in the code size of the optimized binaries MLGOPerf generates, and that is measured as 12\% and 16\% against LLVM's O3 and MLGO on Cbench and 17.8\% and 23.4\% for SPEC CPU2006. MLGOPerf's is trained to optimize performance as its objective and it is reasonable to observe an increase in the code size, especially compared with MLGO which was solely trained to optimize code size.

\begin{table*}[t!]
\centering
\caption{Performance improvement against LLVM's O3 and MLGO}
\label{tab:mlgoperf_results}
\subfloat[SPEC CPU2006 (w/ ref dataset) Performance Improvement]{
    \label{tab:mlgoperf_results_spec}
    \resizebox{!}{!}{%
\begin{tabular}{|l|c|c|c|c|c|c|}
\hline
\textbf{Benchmark} & \textbf{\begin{tabular}[c]{@{}c@{}}Speedup \\ wrt O3\end{tabular}} & \textbf{\begin{tabular}[c]{@{}c@{}}Measured\\  Variance\end{tabular}} & \textbf{\begin{tabular}[c]{@{}c@{}}Speedup\\  wrt MLGO\end{tabular}} & \textbf{\begin{tabular}[c]{@{}c@{}}Measured\\  Variance\end{tabular}} & \textbf{\begin{tabular}[c]{@{}c@{}}Code size\\  Increase wrt O3\end{tabular}} & \textbf{\begin{tabular}[c]{@{}c@{}}Code size\\  Increase wrt MLGO\end{tabular}} \\ \hline
401.bzip2          & 1.052                                                              & 0.966                                                                 & 1.072                                                                & 0.594                                                                 & 1.131                                                                         & 1.248                                                                           \\ \hline
403.gcc            & 1.022                                                              & 0.921                                                                 & 1.054                                                                & 0.004                                                                 & 1.227                                                                         & 1.411                                                                           \\ \hline
429.mcf            & 1.009                                                              & 1.021                                                                 & 1.031                                                                & 1.242                                                                 & 1.047                                                                         & 1.077                                                                           \\ \hline
445.gobmk          & 1.030                                                              & 0.249                                                                 & 1.044                                                                & 0.135                                                                 & 1.097                                                                         & 1.104                                                                           \\ \hline
456.hmmer          & 0.997                                                              & 0.040                                                                 & 1.020                                                                & 0.077                                                                 & 1.227                                                                         & 1.273                                                                           \\ \hline
458.sjeng          & 1.003                                                              & 0.354                                                                 & 1.040                                                                & 0.031                                                                 & 1.318                                                                         & 1.373                                                                           \\ \hline
462.libquantum     & 1.040                                                              & 1.856                                                                 & 1.051                                                                & 0.029                                                                 & 1.257                                                                         & 1.428                                                                           \\ \hline
464.h264ref        & 1.068                                                              & 0.620                                                                 & 1.088                                                                & 0.782                                                                 & 1.389                                                                         & 1.312                                                                           \\ \hline
471.omnetpp        & 1.004                                                              & 1.107                                                                 & 0.999                                                                & 1.091                                                                 & 1.146                                                                         & 1.198                                                                           \\ \hline
433.milc           & 1.021                                                              & 0.566                                                                 & 0.999                                                                & 0.486                                                                 & 1.297                                                                         & 1.276                                                                           \\ \hline
444.namd           & 0.992                                                              & 0.530                                                                 & 1.015                                                                & 0.016                                                                 & 1.002                                                                         & 1.018                                                                           \\ \hline
453.povray         & 0.997                                                              & 0.416                                                                 & 1.035                                                                & 0.022                                                                 & 1.237                                                                         & 1.418                                                                           \\ \hline
470.lbm            & 1.020                                                              & 0.025                                                                 & 1.004                                                                & 0.005                                                                 & 1.025                                                                         & 1.031                                                                           \\ \hline
482.sphinx3        & 0.993                                                              & 0.676                                                                 & 0.992                                                                & 0.070                                                                 & 1.167                                                                         & 1.225                                                                           \\ \hline
\textbf{Geomean}   & \textbf{1.018}                                                     & \textbf{0.434\%}                                                        & \textbf{1.031}                                                       & \textbf{0.086\%}                                                        & \textbf{1.178}                                                                & \textbf{1.235}                                                                  \\ \hline
\end{tabular}}
}
\vspace{0.5cm}
\subfloat[Cbench Performance Improvement]{
    \label{tab:mlgoperf_results_cbench}
    \resizebox{!}{!}{%
\begin{tabular}{|l|c|c|c|c|c|c|}
\hline
\textbf{Benchmark}    & \textbf{\begin{tabular}[c]{@{}c@{}}Speedup \\ wrt O3\end{tabular}} & \textbf{\begin{tabular}[c]{@{}c@{}}Measured\\ Variance\end{tabular}} & \textbf{\begin{tabular}[c]{@{}c@{}}Speedup \\ wrt MLGO\end{tabular}} & \textbf{\begin{tabular}[c]{@{}c@{}}Measured\\  Variance\end{tabular}} & \textbf{\begin{tabular}[c]{@{}c@{}}Code size \\ Increase wrt O3\end{tabular}} & \textbf{\begin{tabular}[c]{@{}c@{}}Code size\\  Increase wrt MLGO\end{tabular}} \\ \hline
automotive\_bitcount  & 1.002                                                              & 0.18\%                                                               & 1.005                                                                & 0.17\%                                                                & 1.000                                                                         & 1.000                                                                           \\ \hline
automotive\_qsort1    & 1.000                                                              & 0.14\%                                                               & 1.003                                                                & 0.12\%                                                                & 1.000                                                                         & 1.000                                                                           \\ \hline
automotive\_susan\_c  & 1.020                                                              & 0.32\%                                                               & 1.122                                                                & 1.02\%                                                                & 1.349                                                                         & 1.567                                                                           \\ \hline
automotive\_susan\_e  & 1.015                                                              & 0.28\%                                                               & 1.044                                                                & 1.71\%                                                                & 1.349                                                                         & 1.567                                                                           \\ \hline
automotive\_susan\_s  & 1.002                                                              & 0.29\%                                                               & 1.008                                                                & 0.23\%                                                                & 1.349                                                                         & 1.567                                                                           \\ \hline
bzip2d                & 1.025                                                              & 0.70\%                                                               & 1.012                                                                & 0.70\%                                                                & 1.173                                                                         & 1.239                                                                           \\ \hline
bzip2e                & 1.014                                                              & 0.53\%                                                               & 1.041                                                                & 0.53\%                                                                & 1.113                                                                         & 1.189                                                                           \\ \hline
consumer\_jpeg\_c     & 1.002                                                              & 0.41\%                                                               & 1.005                                                                & 0.37\%                                                                & 1.045                                                                         & 1.200                                                                           \\ \hline
consumer\_jpeg\_d     & 1.041                                                              & 0.22\%                                                               & 1.335                                                                & 1.36\%                                                                & 1.037                                                                         & 1.198                                                                           \\ \hline
consumer\_lame        & 1.006                                                              & 0.13\%                                                               & 1.007                                                                & 0.10\%                                                                & 1.181                                                                         & 1.137                                                                           \\ \hline
consumer\_tiff2bw     & 1.027                                                              & 0.30\%                                                               & 1.023                                                                & 0.32\%                                                                & 1.298                                                                         & 1.356                                                                           \\ \hline
consumer\_tiff2rgba   & 1.051                                                              & 0.38\%                                                               & 1.110                                                                & 1.71\%                                                                & 1.294                                                                         & 1.351                                                                           \\ \hline
consumer\_tiffdither  & 1.006                                                              & 0.20\%                                                               & 1.014                                                                & 0.11\%                                                                & 1.190                                                                         & 1.146                                                                           \\ \hline
consumer\_tiffmedian  & 1.012                                                              & 0.12\%                                                               & 1.056                                                                & 0.16\%                                                                & 1.048                                                                         & 1.048                                                                           \\ \hline
network\_dijkstra     & 1.005                                                              & 1.20\%                                                               & 0.991                                                                & 1.30\%                                                                & 1.000                                                                         & 1.000                                                                           \\ \hline
network\_patricia     & 1.000                                                              & 0.35\%                                                               & 0.999                                                                & 0.29\%                                                                & 1.007                                                                         & 1.052                                                                           \\ \hline
office\_stringsearch1 & 0.996                                                              & 0.50\%                                                               & 1.000                                                                & 0.91\%                                                                & 1.000                                                                         & 1.000                                                                           \\ \hline
security\_blowfish\_d & 1.005                                                              & 0.75\%                                                               & 1.002                                                                & 1.73\%                                                                & 1.000                                                                         & 1.000                                                                           \\ \hline
security\_blowfish\_e & 1.059                                                              & 0.13\%                                                               & 1.029                                                                & 1.80\%                                                                & 1.412                                                                         & 1.390                                                                           \\ \hline
security\_rijndael\_d & 1.042                                                              & 0.64\%                                                               & 1.039                                                                & 0.21\%                                                                & 1.412                                                                         & 1.390                                                                           \\ \hline
security\_rijndael\_e & 1.037                                                              & 0.07\%                                                               & 1.035                                                                & 0.13\%                                                                & 1.009                                                                         & 1.009                                                                           \\ \hline
security\_sha         & 1.134                                                              & 1.63\%                                                               & 1.132                                                                & 0.09\%                                                                & 1.009                                                                         & 1.009                                                                           \\ \hline
telecom\_adpcm\_c     & 1.001                                                              & 0.13\%                                                               & 1.001                                                                & 0.21\%                                                                & 0.996                                                                         & 0.996                                                                           \\ \hline
telecom\_adpcm\_d     & 1.001                                                              & 0.15\%                                                               & 1.000                                                                & 0.19\%                                                                & 1.000                                                                         & 1.000                                                                           \\ \hline
telecom\_CRC32        & 1.065                                                              & 0.44\%                                                               & 1.066                                                                & 0.57\%                                                                & 1.000                                                                         & 1.000                                                                           \\ \hline
\textbf{Geomean}      & \textbf{1.022}                                                     & \textbf{0.3\%}                                                       & \textbf{1.041}                                                       & \textbf{0.4\%}                                                        & \textbf{1.121}                                                                & \textbf{1.161}                                                                  \\ \hline
\end{tabular}}
    }
\end{table*}
\vspace{1cm}

\subsection{Performance Enablement with Autotuning}
\label{sec:res:autotuning}

\begin{table*}[t!]
\centering
\caption{MLGOPerf Enhanced Autotuning Code Region Opportunities}
\label{tab:mlgoperf_autotuning_results_cbench}
\hspace*{-.25cm}
\begin{tabular}{|l|cc|cc|cccc|}
\hline
\multirow{3}{*}{\textbf{Cbench}} & \multicolumn{2}{c|}{\textbf{O3}}                                                                                                                                                                      & \multicolumn{2}{c|}{\textbf{MLGO}}                                                                                                                                                                  & \multicolumn{4}{c|}{\textbf{MLGOPerf}}                                                                                                                                                                                                                                                                                                                   \\ \cline{2-9} 
                                 & \multicolumn{1}{c|}{\multirow{2}{*}{\textbf{\begin{tabular}[c]{@{}c@{}}Autotuning  \\  Speedup\end{tabular}}}} & \multirow{2}{*}{\textbf{\begin{tabular}[c]{@{}c@{}}Tunable\\  Regions\end{tabular}}} & \multicolumn{1}{c|}{\multirow{2}{*}{\textbf{\begin{tabular}[c]{@{}c@{}}Autotuning\\  Speedup\end{tabular}}}} & \multirow{2}{*}{\textbf{\begin{tabular}[c]{@{}c@{}}Tunable\\  Regions\end{tabular}}} & \multicolumn{1}{c|}{\multirow{2}{*}{\textbf{\begin{tabular}[c]{@{}c@{}}Autotuning\\  Speedup\end{tabular}}}} & \multicolumn{3}{c|}{\textbf{Tunable Regions}}                                                                                                                                                                                             \\ \cline{7-9} 
                                 & \multicolumn{1}{c|}{}                                                                                          &                                                                                      & \multicolumn{1}{c|}{}                                                                                        &                                                                                      & \multicolumn{1}{c|}{}                                                                                        & \multicolumn{1}{c|}{\textbf{\begin{tabular}[c]{@{}c@{}}Tunable \\ Regions\end{tabular}}} & \multicolumn{1}{c|}{\textbf{\begin{tabular}[c]{@{}c@{}}wrt \\ O3\end{tabular}}} & \textbf{\begin{tabular}[c]{@{}c@{}}wrt \\ MLGO\end{tabular}} \\ \hline
automotive\_bitcount             & \multicolumn{1}{c|}{1.027714}                                                                                  & 20                                                                                   & \multicolumn{1}{c|}{1.019832}                                                                                & 20                                                                                   & \multicolumn{1}{c|}{1.02781}                                                                                 & \multicolumn{1}{c|}{20}                                                                  & \multicolumn{1}{c|}{1.000}                                                      & 1.000                                                        \\ \hline
automotive\_qsort1               & \multicolumn{1}{c|}{1.009412}                                                                                  & 32                                                                                   & \multicolumn{1}{c|}{1.008607}                                                                                & 32                                                                                   & \multicolumn{1}{c|}{1.009123}                                                                                & \multicolumn{1}{c|}{32}                                                                  & \multicolumn{1}{c|}{1.000}                                                      & 1.000                                                        \\ \hline
automotive\_susan\_c             & \multicolumn{1}{c|}{1.038951}                                                                                  & 116                                                                                  & \multicolumn{1}{c|}{1.036704}                                                                                & 112                                                                                  & \multicolumn{1}{c|}{1.037121}                                                                                & \multicolumn{1}{c|}{188}                                                                 & \multicolumn{1}{c|}{1.621}                                                      & 1.679                                                        \\ \hline
automotive\_susan\_e             & \multicolumn{1}{c|}{1.031977}                                                                                  & 116                                                                                  & \multicolumn{1}{c|}{1.026087}                                                                                & 112                                                                                  & \multicolumn{1}{c|}{1.033349}                                                                                & \multicolumn{1}{c|}{188}                                                                 & \multicolumn{1}{c|}{1.621}                                                      & 1.679                                                        \\ \hline
automotive\_susan\_s             & \multicolumn{1}{c|}{1.001988}                                                                                  & 116                                                                                  & \multicolumn{1}{c|}{1.065891}                                                                                & 112                                                                                  & \multicolumn{1}{c|}{1.146078}                                                                                & \multicolumn{1}{c|}{188}                                                                 & \multicolumn{1}{c|}{1.621}                                                      & 1.679                                                        \\ \hline
bzip2d                           & \multicolumn{1}{c|}{1.15753}                                                                                   & 637                                                                                  & \multicolumn{1}{c|}{1.100431}                                                                                & 580                                                                                  & \multicolumn{1}{c|}{1.165478}                                                                                & \multicolumn{1}{c|}{747}                                                                 & \multicolumn{1}{c|}{1.173}                                                      & 1.288                                                        \\ \hline
bzip2e                           & \multicolumn{1}{c|}{1.032093}                                                                                  & 637                                                                                  & \multicolumn{1}{c|}{1.026258}                                                                                & 580                                                                                  & \multicolumn{1}{c|}{1.033333}                                                                                & \multicolumn{1}{c|}{747}                                                                 & \multicolumn{1}{c|}{1.173}                                                      & 1.288                                                        \\ \hline
consumer\_jpeg\_c                & \multicolumn{1}{c|}{1.040332}                                                                                  & 1049                                                                                 & \multicolumn{1}{c|}{1.017417}                                                                                & 891                                                                                  & \multicolumn{1}{c|}{1.043764}                                                                                & \multicolumn{1}{c|}{1204}                                                                & \multicolumn{1}{c|}{1.148}                                                      & 1.351                                                        \\ \hline
consumer\_jpeg\_d                & \multicolumn{1}{c|}{1.031342}                                                                                  & 1074                                                                                 & \multicolumn{1}{c|}{1.014804}                                                                                & 885                                                                                  & \multicolumn{1}{c|}{1.017778}                                                                                & \multicolumn{1}{c|}{1148}                                                                & \multicolumn{1}{c|}{1.069}                                                      & 1.297                                                        \\ \hline
consumer\_tiff2bw                & \multicolumn{1}{c|}{1.004812}                                                                                  & 641                                                                                  & \multicolumn{1}{c|}{1.018229}                                                                                & 619                                                                                  & \multicolumn{1}{c|}{1.017452}                                                                                & \multicolumn{1}{c|}{903}                                                                 & \multicolumn{1}{c|}{1.409}                                                      & 1.459                                                        \\ \hline
consumer\_tiff2rgba              & \multicolumn{1}{c|}{1.047902}                                                                                  & 633                                                                                  & \multicolumn{1}{c|}{1.122697}                                                                                & 611                                                                                  & \multicolumn{1}{c|}{1.123338}                                                                                & \multicolumn{1}{c|}{907}                                                                 & \multicolumn{1}{c|}{1.433}                                                      & 1.484                                                        \\ \hline
consumer\_tiffdither             & \multicolumn{1}{c|}{1.012297}                                                                                  & 640                                                                                  & \multicolumn{1}{c|}{1.004719}                                                                                & 614                                                                                  & \multicolumn{1}{c|}{1.007853}                                                                                & \multicolumn{1}{c|}{902}                                                                 & \multicolumn{1}{c|}{1.409}                                                      & 1.469                                                        \\ \hline
consumer\_tiffmedian             & \multicolumn{1}{c|}{0.973255}                                                                                  & 741                                                                                  & \multicolumn{1}{c|}{1.001938}                                                                                & 715                                                                                  & \multicolumn{1}{c|}{0.988845}                                                                                & \multicolumn{1}{c|}{1069}                                                                & \multicolumn{1}{c|}{1.443}                                                      & 1.495                                                        \\ \hline
network\_dijkstra                & \multicolumn{1}{c|}{1.078947}                                                                                  & 13                                                                                   & \multicolumn{1}{c|}{1.087719}                                                                                & 13                                                                                   & \multicolumn{1}{c|}{1.061404}                                                                                & \multicolumn{1}{c|}{22}                                                                  & \multicolumn{1}{c|}{1.692}                                                      & 1.692                                                        \\ \hline
network\_patricia                & \multicolumn{1}{c|}{1.015152}                                                                                  & 12                                                                                   & \multicolumn{1}{c|}{1.015152}                                                                                & 12                                                                                   & \multicolumn{1}{c|}{1.008772}                                                                                & \multicolumn{1}{c|}{12}                                                                  & \multicolumn{1}{c|}{1.000}                                                      & 1.000                                                        \\ \hline
office\_rsynth                   & \multicolumn{1}{c|}{0.998958}                                                                                  & 152                                                                                  & \multicolumn{1}{c|}{1.001032}                                                                                & 147                                                                                  & \multicolumn{1}{c|}{1.018398}                                                                                & \multicolumn{1}{c|}{153}                                                                 & \multicolumn{1}{c|}{1.007}                                                      & 1.041                                                        \\ \hline
security\_blowfish\_d            & \multicolumn{1}{c|}{1.001764}                                                                                  & 18                                                                                   & \multicolumn{1}{c|}{1.000441}                                                                                & 18                                                                                   & \multicolumn{1}{c|}{0.998679}                                                                                & \multicolumn{1}{c|}{18}                                                                  & \multicolumn{1}{c|}{1.000}                                                      & 1.000                                                        \\ \hline
security\_blowfish\_e            & \multicolumn{1}{c|}{1.001314}                                                                                  & 18                                                                                   & \multicolumn{1}{c|}{1.002632}                                                                                & 18                                                                                   & \multicolumn{1}{c|}{1.001314}                                                                                & \multicolumn{1}{c|}{18}                                                                  & \multicolumn{1}{c|}{1.000}                                                      & 1.000                                                        \\ \hline
security\_pgp\_d                 & \multicolumn{1}{c|}{1.019659}                                                                                  & 955                                                                                  & \multicolumn{1}{c|}{1.017919}                                                                                & 929                                                                                  & \multicolumn{1}{c|}{1.036332}                                                                                & \multicolumn{1}{c|}{1317}                                                                & \multicolumn{1}{c|}{1.379}                                                      & 1.418                                                        \\ \hline
security\_pgp\_e                 & \multicolumn{1}{c|}{1.039591}                                                                                  & 955                                                                                  & \multicolumn{1}{c|}{1.038804}                                                                                & 929                                                                                  & \multicolumn{1}{c|}{1.04023}                                                                                 & \multicolumn{1}{c|}{1317}                                                                & \multicolumn{1}{c|}{1.379}                                                      & 1.418                                                        \\ \hline
security\_rijndael\_d            & \multicolumn{1}{c|}{1.040965}                                                                                  & 22                                                                                   & \multicolumn{1}{c|}{1.048175}                                                                                & 22                                                                                   & \multicolumn{1}{c|}{1.04694}                                                                                 & \multicolumn{1}{c|}{25}                                                                  & \multicolumn{1}{c|}{1.136}                                                      & 1.136                                                        \\ \hline
security\_rijndael\_e            & \multicolumn{1}{c|}{1.018811}                                                                                  & 22                                                                                   & \multicolumn{1}{c|}{1.02481}                                                                                 & 22                                                                                   & \multicolumn{1}{c|}{1.029064}                                                                                & \multicolumn{1}{c|}{25}                                                                  & \multicolumn{1}{c|}{1.136}                                                      & 1.136                                                        \\ \hline
security\_sha                    & \multicolumn{1}{c|}{1.009674}                                                                                  & 10                                                                                   & \multicolumn{1}{c|}{1.004434}                                                                                & 10                                                                                   & \multicolumn{1}{c|}{1.101781}                                                                                & \multicolumn{1}{c|}{13}                                                                  & \multicolumn{1}{c|}{1.300}                                                      & 1.300                                                        \\ \hline
telecom\_adpcm\_c                & \multicolumn{1}{c|}{1.006329}                                                                                  & 7                                                                                    & \multicolumn{1}{c|}{1.004211}                                                                                & 7                                                                                    & \multicolumn{1}{c|}{1.002101}                                                                                & \multicolumn{1}{c|}{7}                                                                   & \multicolumn{1}{c|}{1.000}                                                      & 1.000                                                        \\ \hline
telecom\_adpcm\_d                & \multicolumn{1}{c|}{1.039636}                                                                                  & 7                                                                                    & \multicolumn{1}{c|}{1.039636}                                                                                & 7                                                                                    & \multicolumn{1}{c|}{1.038986}                                                                                & \multicolumn{1}{c|}{7}                                                                   & \multicolumn{1}{c|}{1.000}                                                      & 1.000                                                        \\ \hline
telecom\_CRC32                   & \multicolumn{1}{c|}{1.003663}                                                                                  & 4                                                                                    & \multicolumn{1}{c|}{1.001217}                                                                                & 4                                                                                    & \multicolumn{1}{c|}{1.003663}                                                                                & \multicolumn{1}{c|}{5}                                                                   & \multicolumn{1}{c|}{1.250}                                                      & 1.250                                                        \\ \hline
telecom\_gsm                     & \multicolumn{1}{c|}{1.010018}                                                                                  & 115                                                                                  & \multicolumn{1}{c|}{1.009991}                                                                                & 112                                                                                  & \multicolumn{1}{c|}{1.007286}                                                                                & \multicolumn{1}{c|}{118}                                                                 & \multicolumn{1}{c|}{1.026}                                                      & 1.054                                                        \\ \hline
\textbf{Geomean}                 & \multicolumn{1}{c|}{\textbf{1.025198}}                                                                         & \textbf{---}                                                                         & \multicolumn{1}{c|}{\textbf{1.027676}}                                                                       & \textbf{---}                                                                         & \multicolumn{1}{c|}{\textbf{1.037887}}                                                                       & \multicolumn{1}{c|}{\textbf{---}}                                                        & \multicolumn{1}{c|}{\textbf{1.218}}                                             & \textbf{1.260}                                               \\ \hline
\end{tabular}
\end{table*}

As discussed earlier, an added benefit of utilizing an optimized function inlining pass is the enablement of tunable opportunities for subsequent passes which may translate into an increased performance. In this section, we experimentally explore this phenomenon by designing an autotuning methodology to detect whether or not MLGOPerf provides an increased code-region opportunity to the code segments of interest and that if it can lead to finding more optimal configurations.  We leverage our autotuner to explore code-region opportunities for \texttt{loop-unroll} and \texttt{loop-vectorize} by tuning unroll-count $\in \{0, 2, 4, 8\}$ and interleave-count $\in \{1, 2, 4\}$.

Iteratively, our autotuner uses the list of opportunities to determine optimal configurations by using different search techniques. Then, the compiler uses the configurations suggested by the autotuner and generates a new executable file. We run the emitted executable and provide the execution times to the autotuner as feedback which will be leveraged by the autotuner to generate a new set of configurations based on the feedback. We repeat this tuning process until the stop criteria are satisfied. For practicality, we set the iteration number to 120 per benchmark and we showcase the results for Cbench. Similar to the other experimental results in this work, we used a trimmed mean of runs per benchmark at each iteration. The geometric mean of the measured variance between the runs is recorded at 0.164\%. 

Table \ref{tab:mlgoperf_autotuning_results_cbench} presents this evaluation. We report O3, MLGO, and our proposed work in the scenario where we start by tuning the available parameters and record the number of code region opportunities the autotuner finds suitable to tune. Columns \texttt{Autotuning Speedup} represent the best found configuration at the end of the tuning given our budget. The speedup values are all normalized to LLVM's O3 and it reveals the following. (1) MLGOPerf enables enhanced autotuning opportunities for subsequent passes. These values are reported under the final two columns as MLGOPerf provides an increased number of up to 21.8\% and 26\% for tunable code regions against LLVM's O3 and MLGO, respectively. (2) Understandably, not all enhanced opportunities will translate into higher performance values. As the autotuner explores the optimization space, many of the combinations do not lead to any better outcome, however, we experimentally observe that using MLGOPerf, the rate at which the former translates into higher speedup values is higher. These values are 2.5\%, 2.7\%, and 3.7\% for LLVM's O3, MLGO, and MLGOPerf, respectively.

\section{Discussion}
\label{sec:discussion}

MLGOPerf is the first step towards an ML-guided type of compiler optimizations targeting performance and although it provides benefits to the function inlining and its subsequent optimization passes, it is still in its infancy and we are distant from an ideal standalone compiler pass that derives optimal decisions using ML. Function inlining is a unique optimization which can easily slow down the performance of an executable; no inlining or aggressive inlining both can hurt the performance of a running code and there is a fine line to walk through between the two extremes. There are a number of challenges and shortcomings we would like to mention here. 

\subsection{Challenges}
\label{sec:discussion:shortcoming}

\paragraph{Function vs. Global speedup} \major{MLGOPerf uses the predicted \texttt{function} speedup values of IR2Perf as a proxy to guide the training of its RL agent towards an optimal function inlining that benefits the \texttt{global} speedup of program. However, we experimentally notice that on 15\% of the cases in our training data, these two metrics are contracting one another in a sense that an increase in function speedup led to a decrease in global speedup. This is a common challenge of optimizing applications using a finer-grained performance metric for which we eluded to in Section \ref{sec:rl_model_training} regarding performance decoupling. There are several factors involved in this phenomenon, i.e., cache hierarchy, memory-bound workloads, hardware microarchitecture design, etc., and are outside the scope of this work \cite{theodoridis2022inlining}.}

\paragraph{Multi-objective optimization} MLGOPerf attempts to optimize the performance of a code segment with intelligent inlining decisions derived by ML. However, in a number of cases, we experimentally observe that the emitted code size is also increased. \major{Although this is outside the scope of the current work, an enhanced function inlining may also benefit from taking into account both of the objectives in its exploration strategy by identifying the Pareto optimality \cite{silva2021exploring}}. Similar strategies are employed for a number of adjacent problems; i.e, Multicore Embedded Systems \cite{ascia2005dse,silvano2011multicube} and Compiler autotuning \cite{hoste2008cole,Ashouri2013VLIW,CK-ReQuest}.  

\paragraph{Compiler optimization space} It is an inherently difficult problem to identify the optimal set of optimization passes given a code segment. There are a plethora of permutations of optimizations which can be applied to increase the performance of a running code and the problem quickly becomes an NP-hard problem \cite{ashouri2018survey}. Additionally, the optimization space is unbounded as there is no decision boundary for the length of an optimization sequence one can add to the optimization pipeline. Similar to Halting problem \cite{burkholder1987halting}, a general algorithm to solve the selection and the phase-ordering problem for all possible program-input pairs cannot exist.

\paragraph{Compiler performance prediction} Another fundamentally difficult problem with MLGOPerf and any other performance prediction framework is the fact that predicting the CPU time/cycles are architecture and compiler dependent. There are many \texttt{unknown} factors in place to affect the performance of a binary and at the same time, we always, at best, measure a noisy estimate of the performance metrics and for these reasons, predicting the performance or speedup remains a difficult problem.

\paragraph{Software characterization} MLGOPerf leverages our handcrafted features to characterize a segment of a code. However, this process has an unwanted discretization noise which can be reduced by means of finer-grained embedding which is outside the scope of this work. We acknowledge that one of the toughest challenges in the compiler space is the lack of an ideal characterization of software in a way that ML models can be applied and we are no exception here.

\section{Conclusion}
\label{sec:conclusion}

In this paper, we presented MLGOPerf, the first end-to-end framework capable of optimizing performance using ML-Inliner. We experimentally demonstrated that using MLGOPerf, we are able to achieve up to 1.8\% and 2.5\% performance speedup over LLVM's O3 on SPEC CPU 2006 and Cbench while expanding the horizons of code-regions opportunities for subsequent passes up to 26\% which can add further 3.7\% speedup to the emitted binary at the end of the O3 pipeline. Future works will be focused around generalizing a flow for which other compiler optimizations can be tackled in a seamless manner.

\balance
\bibliographystyle{ACM-Reference-Format}
\bibliography{bib}

\end{document}